\documentclass[12pt]{article}
\usepackage{amssymb}
\usepackage{amsmath}
\usepackage{graphicx}
\usepackage{color}
\usepackage{subfigure}
\usepackage[all,arc]{xy}

\newcommand{\R}{{\mathbb R}}
\newcommand{\C}{{\mathbb C}}
\newcommand{\vsigma}{\vec{\sigma}}
\newcommand{\vz}{\vec{z}}
\newcommand{\D}{\mathfrak{D}}

\newcommand{\uq}{\stackrel{\rotatebox{180}{Q}}{ }}
\newcommand{\p}{\mbox{\boldmath$\psi$}}
\newcommand{\tslash}{{\backslash\!\!\!t}}
\newcommand{\Tslash}{{\backslash\!\!\!\!\,T}}
\newcommand{\zslash}{{\backslash\!\!\!z}}
\newcommand{\yslash}{{\backslash\!\!\!y}}
\newcommand{\muslash}{{\backslash\!\!\!\mu}}
\newcommand{\Dslash}{{\backslash\!\!\!\!D}}
\newcommand{\omslash}{{\backslash\!\!\!\omega}}
\newcommand{\xslash}{{\backslash\!\!\!x}}
\newcommand{\F}{{\cal F}}

\newcommand{\pl}{\overleftarrow{\partial}}

\renewcommand{\L}{{\cal L}}
\newcommand{\V}{{\rm Vec}}

\title{\ \\ \ \\ \ \\ \bf\LARGE Instantons on the Taub-NUT Space}
\author{\Large 
Sergey A. Cherkis\thanks{E-mail: cherkis@maths.tcd.ie}\\
\\
\it School of Mathematics and\\
\it Hamilton Mathematics Institute,\\
\it Trinity College Dublin,\  Ireland}

\begin{document}
\begin{titlepage}

\renewcommand{\thepage}{ }
\date{}

\maketitle
\abstract{We present a construction of self-dual Yang-Mills connections on the Taub-NUT space.  We illustrate it by finding explicit expressions for all $SU(2)$ instantons of instanton number one and generic monodromy at infinity.}

\vspace{-5.5in}

\noindent\parbox{\linewidth
}
{
\shortstack{\hspace{-0.5in}
{\em\large Dedicated to the memory of Raphael Siev}\hspace{1.2in}  HMI 09-04\\
\hfill TCDMATH 09-08}}

\end{titlepage}

\tableofcontents

\section{Introduction}

An instanton on a Taub-NUT space is a connection, given by a $u(n)$-valued one-form $i A,$ on an n-dimensional Hermitian bundle ${\cal E}$ over the Taub-NUT space with the curvature two-form $F=dA-i A\wedge A$ satisfying the self-duality condition
\begin{equation}
F=*F.
\end{equation}
Here $*$ denotes the Hodge star operator taking a two-form to its dual. We require the connection A to have finite action $S=\int {\rm tr} F\wedge *F.$

Everywhere outside one point $0$ the Taub-NUT space itself can be thought of as being fibered by a circle $S^1$ over a base ${\mathbb R}^3\backslash 0$.  Choosing $\tau\sim\tau+4\pi$ to be the periodic coordinate on the $S^1$ fiber and $\vec{x}=(x_1,x_2,x_3),$ with $x_1, x_2,$ and $x_3$ coordinates on ${\mathbb R}^3,$ the Taub-NUT metric\footnote{The factor of $\frac{1}{4}$ in the metric is chosen for future convenience and the apparent singularity at the origin of ${\mathbb R}^3$ is merely a coordinate singularity.} is
\begin{equation}
ds^2=\frac{1}{4}\left(\left( l+\frac{1}{|\vec{x}|}\right) d\vec{x}^2+\frac{1}{\left( l+\frac{1}{|\vec{x}|}\right)}(d\tau+\vec{\omega}\cdot d\vec{x})^2\right),
\end{equation}
where $l$ is some fixed parameter determining the asymptotic size of the $S^1$ and $\frac{\partial}{\partial x_i}\frac{1}{|\vec{x}|}=\epsilon_{ijk}\frac{\partial}{\partial x_j} \omega_k.$  This metric degenerates to a flat metric on ${\mathbb R}^4$ as $l\rightarrow 0.$  Its noncompact cycle ${\cal C}: \{(\tau, \vec{x}) | x_1=x_2=0, x_3\geq 0\}$ becomes a plane in this limit.

The Taub-NUT space is equipped with a natural line bundle with a connection $a=\frac{1}{2V}(d\tau+\omega).$  This connection has a self-dual curvature $da.$  As a matter of fact, it has a one parameter family of such line bundles with  the following  Abelian connections
\begin{equation}
a_s=s a=\frac{s}{2}\frac{d\tau+\omega}{V},
\end{equation}
parameterized by $s\in[-l/2,l/2].$  These connections are Abelian instantons, as their curvature is self-dual in the orientation $(\tau, x_1, x_2, x_3)$ and has a  finite action.  Note, that the relation between the left and right ends of this interval is given by tensoring with a line bundle ${\cal L}_l,$  which is trivial  since
\begin{equation}
\int_C d(a_{l/2}-a_{-l/2})=2\pi.
\end{equation}

\subsection{Background}
There has been a lot of work exploring instantons in various backgrounds.  The ADHM original construction \cite{Atiyah:1978ri} provides all instantons on $\R^4$.  Nahm modified this construction in \cite{Nahm:1979yw,NahmCalorons} to provide calorons, i.e. instantons on $\R^3\times S^1.$  Orbifolding the ADHM construction Kronheimer and Nakajima \cite{KN} obtained instantons on ALE spaces.  In \cite{Nekrasov:1998ss} Nekrasov and Schwarz modified the ADHM construction to construct instantons on noncommutative $\R^4.$ All of these constructions have string theory interpretations \cite{DM,Johnson:1996py,Diaconescu:1996rk} and emerge from the sigma model analysis of appropriate D-brane configurations.

Based on these general constructions some explicit solutions at a general position were obtained in \cite{Kraan:1998xe, Lee:1998bb} for a caloron and in  \cite{Bianchi:1995xd}  for instantons on certain ALE spaces.

We would like to point out that in all these cases the underlying space is flat, or it has a useful flat limit. 
Here, we aim to find a general construction for generic\footnote{Some special instanton solutions on the Taub-NUT space were obtained in 
\cite{Pope:1978kj,BoutalebJoutei:1979iz,Kim:2000mg,Etesi:2003ei}.} instantons on an essentially curved space.
In particular, building on the bow formalism introduced in \cite{Cherkis:2008ip} to study the moduli spaces of instantons on the Taub-NUT space, we find expressions for the instanton connection.  As an illustration of our construction we find the explicit general solution for a single instanton on a Taub-NUT space.

\subsection{Instanton Number and Monopole Charges}
A generic self-dual $U(n)$ configuration on the Taub-NUT space possesses two types of topological charges: an instanton number $k_0$ and $n$ monopole charges $m_1, m_2, \ldots , m_n.$  The instanton number as well as the monopole charges are given by integers.  A detailed discussion of various charges of instantons on muti-Taub-NUT spaces and their relation with the corresponding brane configurations appeared recently in \cite{Witten:2009xu}.  Here we define charges in a somewhat different fashion. 

For any given $\vec{x}\in\R$ consider the monodromy $W(\vec{x},\tau)\in U(n)$ satisfying $(\partial_\tau-iA_\tau) W(\vec{x},\tau)=0$ and  $W(\vec{x},0)=1,$ so that the monodromy around the circle $S^1_{\vec{x}}$ is $W(\vec{x}, 4\pi).$  The finite action condition implies that the conjugacy class of $\lim_{|\vec{x}|\rightarrow\infty} W(\vec{x}, 4\pi)$ is well defined and does not depend on the direction in which we approach infinity.  We write the eigenvalues of $\lim_{|\vec{x}|\rightarrow\infty} W(\vec{x}, 4\pi)$ as $$\exp\left(\frac{2\pi i \lambda_1}{l} \right), \exp\left(\frac{2\pi i \lambda_2}{l} \right),\ldots, \exp\left(\frac{2\pi i \lambda_n}{l}\right).$$  Here we restrict our attention to the so called `maximal symmetry breaking' case presuming all $\lambda_j$ are distinct and ordered: $-\frac{l}{2}<\lambda_1<\lambda_2<\ldots<\lambda_n<\frac{l}{2}.$  

Consider a sphere $S_R^2=\{\vec{x} | |\vec{x}|=R\}\in{\mathbb R}^3$ of large radius $R.$  Any point on this sphere determines a $\tau$-circle in the Taub-NUT space, so that the union of all these circles is a squashed three sphere $S^3_R.$  Thus, $S^3_R$ is fibered by circles over the $S^2_R$ and, for a Taub-NUT space, this fibration is the Hopf fibration $S^1\rightarrow S^3_R\rightarrow S^2_R.$ Since the total action is finite there is a gauge transformation on $S^3_R$ such that for large radius $R$ the connection $A$ restricted to $S^3_R$ approaches one with $\tau$-independent components.  Let us write this connection with $\tau$-independent components in the form $A=\hat{A}-\hat{\Phi} \frac{d\tau+\omega}{V}.$  Then, the self-duality condition for $A$ is equivalent \cite{Kronheimer:1985} to the Bogomolny equation 
\begin{equation}
\hat{F}=*_3 D_{\hat{A}}\hat{\Phi},
\end{equation}
for $(\hat{A}, \hat{\Phi}).$  Here $*_3$ is the three-dimensional Hodge star operator for the flat metric $dx_1^2+dx_2^2+dx_3^2$ and $\hat{F}$ is the curvature form of $\hat{A}.$  The asymptotic eigenvalues of $\hat{\Phi}$ are determined by the eigenvalues of the monodromy operator $W(\vec{x}, 4\pi).$ Moreover, since the asymptotic behavior of $\hat{\Phi}$ eigenvalues is the same as for a BPS monopole, the eigenvalues of $\hat{\Phi}$ are
\begin{align}
&\lambda_1+\frac{j_1}{x}+O(x^{-2}), &&\lambda_2+\frac{j_2}{x}+O(x^{-2}), &\ldots\ \ &, &&\lambda_n+\frac{j_n}{x}+O(x^{-2}),
\end{align} 
with $j_1, j_2, \ldots, j_n$ integers.  

Let us describe this construction in different terms, making clear that $j$'s are indeed integers.  Considering the eigen-spaces of the monodromy operator $W(\vec{x}, \tau+4\pi) W^{-1}(\vec{x}, \tau)$ we split the bundle ${\cal E}|_{S^3_R}$ into $k$ eigen-line bundles  ${\cal E}|_{S^3_R}=\L_{\lambda_1}\oplus\L_{\lambda_2}\oplus\ldots\L_{\lambda_n}.$   Since each eigenvalue $\lambda$ is independent of the base, each of these line bundles can be trivialized on all $S^1$ Hopf fibers simultaneously. Thus we have a well defined pushdown line bundles over the base of the Hopf fibration $S^2_R.$  Chern classes of these are $j_1, j_2, \ldots, j_n.$ We now use these integers to define the monopole charges of the configuration.

Let $M=\min(j_1, j_1+j_2,\ldots, j_1+j_2+\ldots+j_n).$ The monopole charges  of an instanton on a Taub-NUT are defined as $$(m_1, m_2, \ldots, m_n)=(j_1-M, j_1+j_2-M,\ldots, j_1+j_2+\ldots+j_n-M).$$  Note, that from the way they are defined, one of these charges, say $m_p,$ must vanish.  Nevertheless, we keep it among the charges and its position $p$ is significant as will be clear momentarily.

Intuitively, since the total action is finite, the asymptotic connection can be put into a form independent of the $\tau$ coordinate.  Then, asymptotically, it can be reduced to a monopole on the base $\R^3$  \cite{Kronheimer:1985}.  It is the charges of this monopole that we defined above.

The instanton number is less straightforward to define.  One can write an explicit expression given by the Chern number minus the contributions of the monopole charges.  To make clear that it is integer, we define it here as an index of the Weyl operator for the connection $A+\frac{1}{2}(\lambda_p+\lambda_{p+1})a$:
\begin{equation}
k_0={\rm Ind}\  \Dslash_{A+\frac{1}{2}(\lambda_p+\lambda_{p+1})a}.
\end{equation}
Thus a general $U(n)$ instanton on a Taub-NUT has an instanton number $k_0$ and monopole charges $(m_1, m_2,\ldots, m_n).$

Kronheimer \cite{Kronheimer:1985} demonstrated equivalence of the `pure monopole' case, i.e. the case with  $k_0=0,$ to singular monopoles studied in \cite{Cherkis:1997aa, Cherkis:1998hi}.  In particular, explicit solutions for $k_0=0$ and $m=1$  (that is $(1,0)$ monopole charges) are equivalent to singular monopole solutions presented in \cite{Cherkis:2007qa,Cherkis:2007jm}. In this paper we focus our attention on the pure instanton case of vanishing monopole charges, and obtain explicit solutions with $k_0=1,$ i.e. a single $SU(2)$   instanton on the Taub-NUT space with no monopole charge.  The explicit metric on the moduli space of such solutions was found in \cite{Cherkis:2008ip}.

\section{Ingredients}
The data specifying an instanton on a Taub-NUT space will be encoded in terms of a {\em bow diagram}.  
There are two basic ingredients in our construction: arrows and strings.
\begin{figure}[htbp]
\begin{center}
\subfigure[Linear maps (arrows and limbs).]
{
\includegraphics[width=0.4\textwidth]{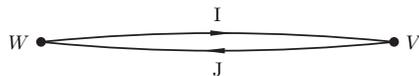} 
}
\hspace{1.5cm}
\subfigure[Nahm Data (string).]
{
\includegraphics[width=0.4\textwidth]{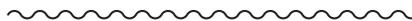}
}
\caption{Components of bow diagrams.}
\label{ingredients}
\end{center}
\end{figure}

\subsection{Arrows and Limbs} 
Figure 1a represents a pair of complex vector spaces $V=\C^v $ and $W=\C^w$ with  maps $J: V\rightarrow W$ and $I: W\rightarrow V.$  The linear space formed by the pair of maps $(I,J)$ has a natural hyperk\"ahler structure, which is respected by the action of $U(v)$ and $U(w).$  The hyperk\"ahler moment map of the $U(v)$ action $g_v:(I,J)\mapsto(g_v^{-1} I, J g_v)$ is given by
\begin{equation}
\mu_V^\C=\mu_V^1+i\mu_V^2=I J,\ \ \ \mu_V^\R=\mu_V^3=\frac{1}{2}(J^\dagger J-I I^\dagger),
\end{equation}
while for the $U(w)$ action $g_w: (I,J)\mapsto ( I g_w, g_w^{-1} J )$ the moment map is
\begin{equation}
\mu_W^\C=\mu_W^1+i\mu_W^2=-J I,\ \ \ \mu_W^\R=\mu_W^3=\frac{1}{2}(I^\dagger I-J J^\dagger).
\end{equation}
It is convenient to assemble the pair $(I,J)$ into 
\begin{equation}
Q=\left(\begin{array}{c} J^\dagger\\ I\end{array}\right)\ \mathrm{and}\ 
\uq=\left(\begin{array}{c} I^\dagger\\ -J\end{array}\right),
\end{equation}
(pronounced ``kyu'' and ``yuk'') so that $Q: W\rightarrow S\otimes V$ and $\uq: V\rightarrow S\otimes W$ 
with the three complex structures $e_j=-i\sigma_j$ acting on $Q$'s.  $S\approx {\mathbb C}^2$ is a two dimensional space of spinors providing the representation of quaternions, with the quaternionic units
 $e_j=i\sigma_j,$ i.e. 
$$
e_1=-i
\Bigl(
\begin{array}{cc}
\scriptstyle 0 & \scriptstyle 1 \\
 \scriptstyle 1 & \scriptstyle 0   
\end{array}
\Bigr),\ 
e_2=-i
\Bigl(
\begin{array}{cc}
\scriptstyle 0 & \scriptstyle -i \\
\scriptstyle i & \scriptstyle 0   
\end{array}
\Bigr),\ 
e_3=-i
\Bigl(
\begin{array}{cc}
\scriptstyle 1 & \scriptstyle 0 \\
\scriptstyle 0 & \scriptstyle -1   
\end{array}
\Bigr).
$$
The natural metric on the linear space of all pairs of maps is 
\begin{equation}
ds^2={\rm tr}_W dQ^\dagger dQ={\rm tr}_W(dJ dJ^\dagger+dI^\dagger dI).
\end{equation} 
The tree symplectic forms 
\begin{equation}
\omega_j\equiv g(\cdot, e_j\cdot)=\frac{1}{2}{\rm tr}_W(dQ^\dagger\wedge e_j dQ),
\end{equation} 
can be combined into $\omslash\equiv\omega_j\sigma_j= i \V\,{\rm tr}_V\, dQ\wedge dQ^\dagger.$  Here we introduce a `vector operation' $\V$ defined by 
\begin{equation}
\V(1_{2\times 2}\otimes M^0 +\sigma_j\otimes M^j)=\sigma_j \otimes M^j.
\end{equation}
Since $-i\sigma_j$ represent the quaternionic imaginary units, this operation basically amounts to taking the imaginary part of a quaternion.

With this notation the moment maps are 
\begin{equation}
\muslash_V=\mu_V^i \sigma_i=\V(Q Q^\dagger)\ \ \mathrm{and}\ \ 
\muslash_W=\mu_W^i \sigma_i=\V(\uq \uq^\dagger).
\end{equation}

\subsection{The String}
Figure 1b represents an interval $\cal I$ parameterised by $s$ with a  bundle $E\rightarrow\cal I$ endowed with a Hermitian structure, a connection $D_s=d/ds+iT_0,$ and a triplet $\vec{T}=(T_1,T_2,T_3)$ of endomorphisms of $E.$   In other words for a given trivialization of $E$ we have a quadruplet of Hermitian matrix valued functions $(T_0(s), T_1(s), T_2(s), T_3(s)).$  These also form a linear space with a natural flat metric
$ds^2=\int{\rm tr}_E\left(dT_0^2+dT_1^2+dT_2^2+dT_3^2\right)$ 
and  a hyperk\"ahler structure all invariant with respect to the following gauge group action
\begin{equation}
g(s): \left(\begin{array}{c}T_0(s)\\ T_1(s)\\ T_2(s)\\ T_3(s)\end{array}\right)\mapsto
\left(\begin{array}{c}g^{-1} T_0 g-i g^{-1}\frac{d}{ds}g\\ g^{-1}T_1 g\\ g^{-1}T_2 g\\ g^{-1}T_3 g\end{array}\right). 
\end{equation}
The corresponding moment maps are
\begin{eqnarray}
\mu^1&=&\frac{d}{ds}T_1+i[T_0,T_1]+i[T_2,T_3],\\
\mu^2&=&\frac{d}{ds}T_2+i[T_0,T_2]+i[T_3,T_1],\\
\mu^3&=&\frac{d}{ds}T_3+i[T_0,T_3]+i[T_1,T_2].
\end{eqnarray}
It is convenient to introduce $\Tslash=\sigma_1\otimes T_1+\sigma_2\otimes T_2+\sigma_3\otimes T_3$ so that the moment map $\muslash=[\frac{d}{ds}+i T_0, \Tslash]+\V\,\Tslash \Tslash.$

Assembling the Nahm data into a quaternion $T=T_0+T_j\otimes e_j=T_0-i\Tslash$ we write the above metric  on the linear space of all the Nahm data in the form
\begin{equation}
ds^2=\frac{1}{2}\int {\rm tr}_S\, {\rm tr}_E\, \delta T^\dagger\delta T ds.
\end{equation}
The three symplectic forms $\omega_j=g(\cdot, e_j\cdot)$ are encoded in 
\begin{equation}
\omslash=\frac{i}{2}\int {\rm tr}_E \delta T\wedge\delta T^\dagger ds.
\end{equation}
Note that the moment maps can be written in terms of the Weyl operator $\Dslash=-D_s+\Tslash$ and its conjugate $\Dslash^\dagger=D_s+\Tslash$ as
\begin{equation}
\muslash=\V(D_s+\Tslash)(-D_s+\Tslash).
\end{equation}

\section{The Taub-NUT as a Hyperk\"ahler Quotient}
This section contains a description of the Taub-NUT space using the ingredients we have defined in the previous section.  This description will  naturally lead us to a family  of self-dual harmonic forms\footnote{A description of these in terms of the hyperk\"ahler reduction recently appeared in \cite{Witten:2009xu}.} which are essential for the instanton construction that follows.  Our exposition in this section is close to that of Gibbons and Rychenkova \cite{Gibbons:1996nt}.
Just as for the construction \cite{KN} of instantons on ALE spaces it was essential to know the realization of the underlying ALE space as a hyperk\"ahler quotient of linear spaces \cite{Kronheimer}, this section contains the realization of the Taub-NUT space as a hyperk\"ahler quotient setting the groundwork for the construction of instantons on it.

\subsection{Taub-NUT Bow Data}
\begin{figure}
\begin{center}
\includegraphics[width=0.6\textwidth]{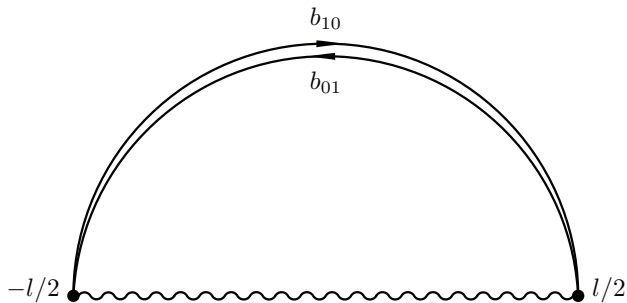}
\caption{A Taub-NUT Bow Diagram.}
\label{fig:TN}
\end{center}
\end{figure}

The bow diagram on Figure \ref{fig:TN} represents Nahm data of rank $1$ assosiated with a Hermitian line bundle $e\rightarrow\cal I$ on an interval $[-l/2,l/2]$ of length $l,$ as well as maps $b_{10}$ and $b_{01}$ between the one-dimensional complex vector spaces $e_0=e|_{s=-l/2}$ and $e_1=e|_{s=l/2}$ at the ends of the interval.  
A gauge transformation $h(s)$ acts on these data as follows:
\begin{equation}
\left(\begin{array}{c} t_0 \\ t_j \\ b_{01} \\ b_{10} \end{array}\right)\mapsto
\left(\begin{array}{c} h^{-1}t_0 h + i h^{-1}\frac{d}{ds} h \\ h^{-1}t_j h \\ h^{-1}(-{\scriptstyle \frac{l}{2}}) b_{01} h({\scriptstyle \frac{l}{2}}) \\ h^{-1}({\scriptstyle \frac{l}{2}}) b_{10} h(-{\scriptstyle \frac{l}{2}}) \end{array}\right)
\end{equation}
Introducing ${\bf t}=t_1+i t_2 $ and ${\bf\scriptstyle D}=d/ds-i t_0-t_3$ the vanishing of moment maps can be written in complex notation as
\begin{eqnarray}\label{eq:TNmom}
&&[{\bf\scriptstyle D}, {\bf t}]-\delta{\scriptstyle (s+\frac{l}{2})} b_{01} b_{10}+\delta{\scriptstyle (s-\frac{l}{2})}b_{10} b_{01}=0,\\
&&[{\bf\scriptstyle D}^\dagger, {\bf\scriptstyle D}]+[{\bf t}^\dagger, {\bf t}]+\delta{\scriptstyle (s+\frac{l}{2})}(b_{10}^\dagger b_{10}-b_{01} b_{01}^\dagger)+\delta{\scriptstyle (s- \frac{l}{2})}(b_{01}^\dagger b_{01}-b_{10} b_{10}^\dagger)=0.\nonumber
\end{eqnarray}

Let us distinguish some point $s_0$ on the Nahm interval. Say this point divides this interval into two intervals of lengths $l_L$ and $l_R,$ i.e. $l_L+l_R=l$ and at this distinguished point $s=s_0=l_L-l/2=l/2-l_R.$ Let us assume $s_0>0.$ We shall perform the hyperk\"ahler quotient step-by-step, so that the last step is the quotient with respect to the $U(1)$ at the distinguished point\footnote{To be exact, this $U(1)$ is the quotient of the group of all gauge transformation on the interval by the subgroup formed by the gauge transformation that equal to identity at $s_0.$} $s_0.$ This will allow us to associate with any point $s_0$ a natural line bundle over the Taub-NUT space and its natural connection corresponding to this $U(1).$

\subsection{The Family of Connections}
First we perform hyperk\"ahler reduction on each open interval separately.  The intervals are of lengths $l_L$ and $l_R.$ Since the computations are identical,  we focus on the interval of length $l_R$ to the right of $s_0.$  As the Nahm data is Abelian, the vanishing of the moment maps implies $d t_j/ds=0,$ thus the vector $\vec{t}=(t_1, t_2, t_3)$ is constant.  The connection $t_0$ can be made constant using gauge transformations that are trivial at the ends of the interval.  There is a large gauge transformation $g=\exp(2\pi i (s-s_0)/l_R)$ satisfying $g(s_0)=g(l/2)=1.$ This gauge transformation takes $t_{0R}$ to $t_{0R}+2\pi/l_R.$ Thus the result of this hyperk\"ahler reduction is $\R^3\times S^1$ with coordinates $t_{1R},t_{2R},t_{3R}$ and $t_{0R}\sim t_{0R}+2\pi/l_R$ and the metric
\begin{equation}
ds^2=\int_{s_0}^l \left( dt_{0R}^2+d{\vec{t}_R}^{\ 2}\right) ds=l_R\left(dt_{0R}^2+d\vec{t}_R^{\ 2}\right).
\end{equation}
The resulting metric on the Nahm data on the left interval is given by the same expression with $l_R$ replaced by $l_L.$

Now we perform the hyperk\"ahler reduction with respect to the $U(1)$ at $s=l/2,$ which can be realized by $h=\exp(i\phi\frac{s-s_0}{l_R}).$  Exploiting the fact that all the data is Abelian, we assemble the linear data $(b_{01}, b_{10})$ into a quaternion 
\begin{equation}
q=q^0+q^i e_i=\left( b_-, b_+\right)=\left(\begin{array}{cc}
\bar{b}_{01} & \bar{b}_{10}  \\-b_{10}& b_{01}
\end{array}\right).
\end{equation}
Here  $b_-$ and $b_+$ play the roles of  $\uq$ and $Q.$ The natural metric is 
\begin{equation}
ds^2=\frac{1}{2}{\rm tr}_S dq^\dagger dq=db^\dagger_- db_-=d b^\dagger_+ db_+,
\end{equation}
and the resulting symplectic forms are given by $\omslash=i\V\, dq\wedge dq^\dagger.$
A gauge transformation $h(s)$ with  $h(-l/2)=\exp(i\phi_L)$ and $h(l/2)=\exp(i\phi_R)$ sends $q$ to $q\exp\big(e_3(\phi_R-\phi_L)\big).$  The resulting moment maps are $\muslash_L=-\frac{1}{2}q\sigma_3 q^\dagger$ and $\muslash_R=\frac{1}{2}q\sigma_3 q^\dagger.$

The rightmost $U(1)$ acts as 
\begin{equation}
\exp({i\phi\frac{s-s_0}{l_R}}): (q,t_{0R}, \vec{t}_R)\mapsto(q e^{e_3 \phi}, t_{0R}-\phi/l_R, \vec{t}_R),
\end{equation} 
with the moment map $\mu_1 e_1+\mu_2 e_2+\mu_3 e_3=\frac{1}{2}q e_3 \bar{q}=t_1 e_1+t_2 e_2+t_3 e_3.$  Let $q=a e^{e_3 \psi/2}$ where $a$ is a pure imaginary quaternion, and let $\vec{x}=(x_1, x_2, x_3)$ be such that $x_1 e_1+x_2 e_2+x_3 e_3=q e_3 \bar{q}.$  The periodic coordinate $\psi\sim\psi+4\pi$ and the components of $\vec{x}$ provide new coordinates on $\R^4$. Then the flat metric on the set of octuplets $(t_{0R}, \vec{t}_{0R}, b_{01}, b_{10})$ is
\begin{eqnarray}
ds^2&=&\frac{1}{2}{\rm tr}_S dq^\dagger dq+l_R\bigl(
{dt}_{0R}^2+{d\vec{t}_R}^2\bigr)\\ &=&
\frac{1}{4}\left(\frac{1}{|\vec{x}|}d\vec{x}^2+|\vec{x}| (d\psi+\omega)^2\right)+l_R\left({dt}_{0R}^2+{d\vec{t}_R}^2\right),
\end{eqnarray}
where 
\begin{equation}\label{Eq:b-relations}
i |\vec{x}| (\omega+d\psi)=\frac{i}{2}{\rm tr}\big(q e_3 dq^\dagger-dq e_3 q^\dagger\big)=db_-^\dagger b_- -b_-^\dagger db_-= -db_+^\dagger b_+ + b_+^\dagger db_+.
\end{equation}
One can easily verify that $\omega=\omega_j dx_j$ satisfies $\epsilon_{ijk}\partial_j\omega_k=\partial_i \frac{1}{|\vec{x}|}.$

The $U(1)$ is acting by $e^{i\phi}: (\psi, t_{0R})\mapsto(\psi+2\phi, t_{0R}-\phi/l_R).$  The invariant of this action is $\sigma=\psi+2 l_R t_{0R}$ and the vanishing of the moment maps implies $\vec{t}_R=-\frac{1}{2}\vec{x}.$ One can readily verify that the above metric becomes
\begin{equation}
ds^2=\frac{1}{4}\left(\bigg(l_R+\frac{1}{|\vec{x}|}\bigg) d\vec{x}^2+\frac{(d\sigma+\omega)^2}{l_R+1/|\vec{x}|}\right)+l_R r\bigg(l_R+\frac{1}{|\vec{x}|}\bigg)\left(dt_{0R}+\frac{1}{2} \frac{d\sigma+\omega}{l_R+1/|\vec{x}|}\right)^2.
\end{equation}
After factoring out the $e^{i\phi}$ action the result is 
\begin{equation}
ds^2=\frac{1}{4}\left(\bigg(l_R+\frac{1}{|\vec{x}|}\bigg) d\vec{x}^2+\frac{(d\sigma+\omega)^2}{l_R+1/|\vec{x}|}\right).
\end{equation}

The last step in the hyperk\"ahler reduction procedure is the hyperk\"ahler quotient with respect to the $U(1)$ at the distinguished point $s=s_0.$  In order to represent this action we use the gauge transformation
\begin{equation}
h(s)=\begin{cases}\exp\big(i\frac{s}{s_0}\varepsilon\big)& {\rm for}\  s\leq s_0\\
\exp\Big(i\frac{l/2-s}{l/2-s_0}\varepsilon\Big)& {\rm for}\  s>s_0\end{cases},
\end{equation}
that is continuous and equals identity at $s=0$ and at $s=l/2.$
At $s=s_0$ this gauge transformation is $h(s_0)=e^{i\varepsilon}.$ It has the following action
\begin{equation}
h(s):\left(\begin{array}{c} t_{0L}\\ \vec{t}_L\\ t_{0R}\\ \vec{t}_R \\q\end{array}\right)
\mapsto
\left(\begin{array}{c} t_{0L}-\varepsilon/s_0\\ \vec{t}_L\\ t_{0R}+\varepsilon/l_R\\ \vec{t}_R \\q\exp\Big(e_3\frac{l}{2 s_0}\varepsilon\Big)\end{array}\right).
\end{equation}
The corresponding moment map is $\muslash=\frac{l_L}{s_0}\tslash_L-\tslash_R+\frac{l}{2s_0}\frac{1}{2}\xslash.$ Since the vanishing of the moment maps of the first stage of the reduction implies $\tslash_R=-\frac{1}{2}\xslash,$ it follows that $\muslash=\frac{l_L}{s_0}\big(\tslash_L+\frac{1}{2}\xslash\big).$  Putting $\muslash$ equal to zero we have $\tslash_L=-\frac{1}{2}\xslash$ as well, so $\vec{t}$ is constant on ${\cal I}.$

So far, including the data on the left interval, we have the metric 
\begin{equation}
ds^2=\frac{1}{4}\left(\bigg(l_R+\frac{1}{|\vec{x}|}\bigg) d\vec{x}^2+\frac{(d\sigma+\omega)^2}{l_R+1/|\vec{x}|}\right)+l_L\left(dt_{0L}^2+d\vec{t}_L^2\right).
\end{equation}
Under the above gauge transformation the angle $\sigma=\psi+2 l_R t_{0R}\mapsto \sigma+2\frac{l_L}{s_0}\varepsilon.$ The invariant coordinate is $\tau=\sigma-2l_L t_{0L}=\psi+2 l_R t_{0R}-2 l_L t_{0L},$ and we choose $\varepsilon\sim\varepsilon+2\pi$ instead of $\sigma$ as a coordinate along the circle of the gauge transformation.  
In these coordinates the above metric can be rewritten as
\begin{multline}\label{eq:hkq}
ds^2=\frac{1}{4}\left[\bigg(l+\frac{1}{|\vec{x}|}\bigg)d\vec{x}^2+\frac{1}{l+1/|\vec{x}|}(d\tau+\omega)^2\right]\\
+\frac{l_L \Big(l+\frac{1}{|\vec{x}|}\Big)}{s_0^2\big(l_R+1/|\vec{x}|\big)}\left(d\varepsilon+\frac{s_0}{2}\frac{d\tau+\omega}{\big(l+1/|\vec{x}|\big)}\right)^2.
\end{multline}
The first part of the expression (\ref{eq:hkq}) is the resulting hyperk\"ahler metric of the Taub-NUT space
\begin{equation}
4\, ds_{TN}^2=\bigg(l+\frac{1}{|\vec{x}|}\bigg)d\vec{r}^2+\frac{1}{l+1/|\vec{x}|}(d\tau+\omega)^2,
\end{equation}
here the one-form $\omega$ satisfies $d\omega=*_3 d \frac{1}{|\vec{x}|}.$
The second part of the expression in Eq.~(\ref{eq:hkq}) provides the natural connection $D=d+i s_0 a$ with the one form $s_0 a,$ where
\begin{equation}
a=\frac{1}{2} \frac{d\tau+\omega}{l+\frac{1}{|\vec{x}|}}.
\end{equation}

\subsection{A Basis of Self-dual Two-forms}
Let $V=l+1/|\vec{x}|,$ so that $a=\frac{d\tau+\omega}{2V}.$ Here we observe the following relation 
\begin{equation}\label{Eq:SDForms}
\left(\frac{1}{2}d\xslash-i a\right)^\dagger\wedge\left(\frac{1}{2}d\xslash-i a\right)=\frac{i}{2}\sigma_k\left(\frac{d\tau+\omega}{V}\wedge dx^k+\frac{1}{2}\epsilon_{ijk} dx^i dx^J\right).
\end{equation}
The components of the right-hand-side are self-dual two-forms in the orientation $(\tau, x^1, x^2, x^3)$ providing a basis of self-dual two-forms on the Taub-NUT.  Let us note for future use that since $\frac{1}{2} d\xslash-ia=-(d\tslash+ia),$ in terms of the $\tau$ and $\vec{t}$ coordinates the combination $(d\tslash+ia)^\dagger\wedge(dt+ia)$ is self-dual.

\section{Instanton Data}
\begin{figure}[htbp]
\begin{center}
\includegraphics[width=0.6\textwidth]{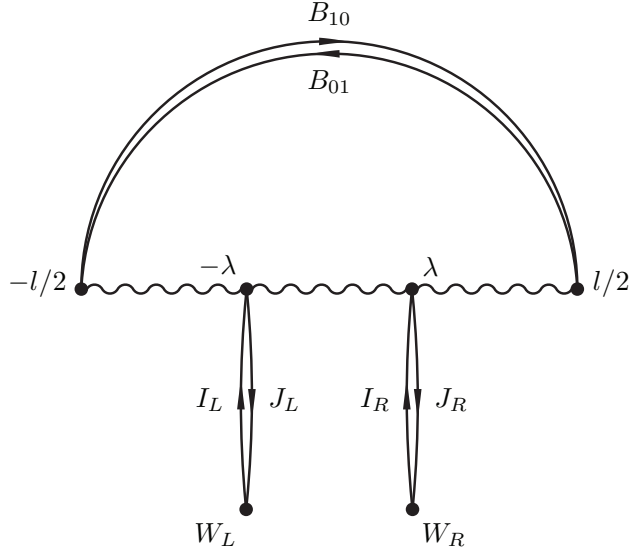}
\caption{The bow diagram for an $SU(2)$ Instanton on the Taub-NUT.}
\label{fig:Instanton}
\end{center}
\end{figure}

Instanton data for an $SU(2)$ instanton with no monopole charges is represented by the bow diagram in Figure \ref{fig:Instanton}. 
It consists of 
\begin{itemize}
\item a rank $k_0$ vector bundle $E\rightarrow[-l/2,l/2]$ with the Nahm data $(T_0, \vec{T})$ on the intervals $[-l/2,-\lambda], [-\lambda,\lambda],$ and $[\lambda, l/2]$ (we do not presume two-sided continuity at $s=\pm\lambda$ across different intervals),  
\item linear maps $B_{10}: E_{-l/2}\rightarrow E_{l/2}$ and $B_{01}: E_{l/2}\rightarrow E_{-l/2},$
\item linear maps 
$I_L: W_L\rightarrow E_{-\lambda},\ 
J_L: E_{-\lambda}\rightarrow W_L,\ 
I_R: W_R\rightarrow E_\lambda,$ and 
$J_R:E_\lambda\rightarrow W_R.$
\end{itemize}
The group of gauge transformations acts on these data as follows
\begin{equation}
g: \left(\begin{array}{c} T_0 \\ T_j \\ B_{01} \\ B_{10} \\  I_{\alpha} \\ J_{\alpha} \end{array}\right)\mapsto
 \left(\begin{array}{c} g^{-1}{\scriptstyle (s)}T_0 g{\scriptstyle (s)} -i g^{-1}{\scriptstyle (s)}\frac{d}{ds} g{\scriptstyle (s)} \\ g^{-1}{\scriptstyle (s)}T_j g{\scriptstyle (s)} \\g^{-1}{(-\scriptstyle \frac{l}{2})} B_{01} g({\scriptstyle \frac{l}{2}}) \\ g^{-1}({\scriptstyle \frac{l}{2}}) B_{10} g(-{\scriptstyle \frac{l}{2}}) \\    g^{-1}{\scriptstyle( \lambda}_{\alpha}{\scriptstyle)} I_{\alpha} \\ J_{\alpha} g{\scriptstyle (\lambda}_\alpha{\scriptstyle)} 
\end{array}\right),
\end{equation}
where the index $\alpha$ takes values $L$ and $R$ and we introduced $\lambda_L=-\lambda$ and $\lambda_R=\lambda.$

Introducing the complex notation $D=\frac{d}{ds}+iT_0-T_3$ and $T=T_1+i T_2,$ the moment maps are written as
\begin{align}\label{Eq:InstMom}
&&[D, T]  +  \delta{\scriptstyle (s+\frac{l}{2})} B_{01} B_{10}  - \delta{\scriptstyle (s-\frac{l}{2})} B_{10} B_{01} + \sum_{\alpha\in\{L,R\}} \delta{\scriptstyle(s-\lambda}_\alpha{\scriptstyle)} I_\alpha J_\alpha=0,\\
&&[D^\dagger, D]+[T^\dagger, T]  +\delta{\scriptstyle (s+\frac{l}{2})}(B_{10}^\dagger B_{10}- B_{01} B_{01}^\dagger)+\delta{\scriptstyle (s-\frac{l}{2})}(B_{01}^\dagger B_{01}-B_{10} B_{10}^\dagger)+\nonumber\\
&&+\sum_{\alpha\in\{L,R\}} \delta{\scriptstyle(s-\lambda}_\alpha{\scriptstyle)} (J_\alpha^\dagger J_\alpha-I_\alpha I^\dagger_\alpha)=0.\nonumber
\end{align}
These conditions can be written compactly if we introduce 
\begin{equation}
B_-=\left(\begin{array}{c}
B_{10}^\dagger \\ -B_{01}\end{array}\right),\ 
B_+=\left(\begin{array}{c}
B_{01}^\dagger \\ B_{10}\end{array}\right),
\end{equation}
and $\Dslash=-\frac{d}{ds}-i T_0+\Tslash=
  \bigl(\begin{smallmatrix}
 -D&T^\dagger\\ T&D^\dagger
  \end{smallmatrix}\bigr).$ Then the moment maps are given by
\begin{equation}
\muslash=\V\left(\Dslash^\dagger \Dslash+\sum_\alpha\delta(s-\lambda_\alpha)Q_\alpha Q^\dagger_\alpha+\delta(s+\frac{l}{2}) B_- B_-^\dagger+\delta(s-\frac{l}{2}) B_+ B_+^\dagger\right).
\end{equation}

\section{The Nahm Transform}
\subsection{The Weyl Operator}
A central role in the ADHM-Nahm transform \cite{Atiyah:1978ri,Nahm:1979yw} is played by a certain linear operator.  In the case at hand it is a modification of the Weyl operator.  The details of similar construction can be found in \cite{NahmCalorons,Hurtubise:1989qy} for the case of calorons. 

Let $H$ be the space of $L^2$ sections of $S\otimes E$ that are continuous on ${\cal I}$ and have $L^2$ derivatives on ${\cal I}\backslash\{\lambda_L, \lambda_R\}.$ Let $\tilde{\cal H}$ be the direct sum of the space of $L^2$ sections of $S\otimes E$ with spaces $W_L, W_R, E_{-l/2},$ and $E_{l/2}.$ Given the instanton data of the bow diagram in Figure \ref{fig:Instanton} we introduce the operator $\D: {\cal H}\to\tilde{\cal H}$ acting by
\begin{equation}
\D: f\mapsto \left(
\begin{array}{c}
\bigl(-\frac{d}{ds}-i T_0+\Tslash\bigr)f\\
(J_L, I^\dagger_L) f(-\lambda)\\
(J_R, I^\dagger_R) f(\lambda)\\
  \bigr(B_{01}, B^\dagger_{10}\bigr) f(l/2)\\
  \bigl(-B_{10}, B^\dagger_{01}\bigr)  f(-l/2)
\end{array}
\right).
\end{equation}
Let us denote by $\psi$ an $L^2$ section of the restriction of $S\otimes E$  to ${\cal I}\backslash\{\lambda_L, \lambda_R\},$ $\chi_\alpha\in E_{\lambda_\alpha},$  $v_{-}\in E_{-l/2}$ and $v_{+}\in E_{l/2}.$
Integrating by parts we find that the cokernel of $\D$ is given by $(\psi(s), \chi_L, \chi_R, v_{-}, v_{+})\in\tilde{\cal H}$  satisfying
\begin{align}
&\left(\frac{d}{ds}+iT_0+\Tslash\right)\psi=0,\ \mathrm{on}\ {\cal I}\backslash\{\alpha_L, \alpha_R\},\\
&\psi(\lambda_\alpha +)-\psi(\lambda_\alpha -)=-Q_\alpha \chi_\alpha,\\
&\psi(l/2)\phantom{-}=\left(\begin{array}{c} B^\dagger_{01}\\ B_{10}\end{array}\right) v_{-},\\
&\psi(-l/2)=-\left(\begin{array}{c} -B^\dagger_{10}\\ B_{01}\end{array}\right)v_{+}.
\end{align}
In other words the dual operator takes the form
\begin{equation}
\begin{split}
\D^\dagger&=\left(\begin{array}{cc} 
-D^\dagger & T^\dagger \\ 
T & D
\end{array}\right)
\oplus\Biggl(\mathop{\oplus}_{\scriptscriptstyle\alpha\in\{L,R\}} \delta{\scriptstyle(s-\lambda}_\alpha\scriptstyle{)}\left(\begin{array}{c} J_\alpha^\dagger \\ I_\alpha\end{array}\right)\Biggr)\\
&\oplus
\left(\delta{\scriptstyle (s+\frac{l}{2})}
\left(\begin{array}{c}
B_{10}^\dagger \\ -B_{01}\end{array}\right)
,\ \delta{\scriptstyle (s-\frac{l}{2})}
\left(\begin{array}{c}
B_{01}^\dagger \\ B_{10}\end{array}\right)
\right),\\
&=\Dslash^\dagger\oplus\delta(s-\lambda_\alpha)Q_\alpha\oplus\Big(\delta({\scriptstyle s+\frac{l}{2}}) B_- , \delta({\scriptstyle s-\frac{l}{2}}) B_+\Big).
\end{split}
\end{equation}
In terms of $\D$ and $\D^\dagger$  the moment map conditions of Eqs.~(\ref{Eq:InstMom}) can be written as 
\begin{equation}
\V (\D^\dagger \D)=0.
\end{equation}

For a given point of the Taub-NUT space of Figure \ref{fig:TN}, corresponding to $(t_0,\vec{t}, b_{10}, b_{01})$ satisfying Eqs.(\ref{eq:TNmom}),  we can twist the above operator  as follows
\begin{equation}\label{Twist}
\begin{split}
\D_t^\dagger&=
\left(\begin{array}{cc} 
-D^\dagger-t_3 & T^\dagger-t^\dagger \\ 
T-t & D+t_3
\end{array}\right)
\oplus\Biggl(\mathop{\oplus}_{\scriptscriptstyle\alpha\in\{L,R\}} \delta(s-\lambda_\alpha)\left(\begin{array}{c}J_\alpha^\dagger \\ I_\alpha\end{array}\right)\Biggr)\\
&\quad \oplus
\left(\delta{\scriptstyle (s+\frac{l}{2})}\left(\begin{array}{cc}B_{10}^\dagger & -b^\dagger_{10}\\ -B_{01} & -b_{01}\end{array}\right)
+\delta{\scriptstyle (s-\frac{l}{2})}\left(\begin{array}{cc} -b^\dagger_{01} & B_{01}^\dagger \\ b_{10}&B_{10}\end{array}\right)\right).
\end{split}
\end{equation}
To be exact, whenever adding two operators with one of them belonging to the instanton bow and another to the Taub-NUT bow data we understand both operators to be tensored with identity so that they act on the tensor product of the corresponding spaces. For example, $T-t$ stands as a shorthand  for $T\otimes 1-1\otimes t.$  Unfortunately, in this case using the rigorous notation would make the formula above much harder to read.  We also allow this shorthand since for a case of a  single instanton the vector spaces are one-dimensional and the bow data is Abelian, so, conveniently, the expression in Eq.~(\ref{Twist}) makes perfect sense as it is written.

\subsection{The Connection}
From now on we understand $\psi$ to be a section of $\C^2\otimes E\otimes e \rightarrow {\cal I}\backslash\{-\lambda,\lambda\},$  $v_{-}\in E_{-l/2}\otimes e_{l/2}$ and $v_{+}\in E_{l/2}\otimes e_{-l/2}.$ We combine $v_+$ and $v_-$ into a spinor $v=\bigl(\begin{smallmatrix}
v_+\\ v_-
\end{smallmatrix}\bigr)$ and  denote the  data $(\psi(s),\chi_L,\chi_R,v)$ by $\p.$  The twisted operator $\D_t^\dagger$ acts on the linear Hermitian space formed by such data.
  For $$\p_1=(\psi_1(s),\chi_{L1},\chi_{R1}^+,v_1) \text{and} \p_2=(\psi_2(s),\chi_{L2},\chi_{R2},v_2)$$ the natural Hermitian product is given by $(\p_1,\p_2)=v_1^\dagger v_2+(\chi_{L1})^\dagger \chi_{L2}+(\chi_{R1})^\dagger \chi_{R2}+\int_{-l/2}^{l/2} \psi_1^\dagger(s) \psi_2(s) ds.$  We also define the operator ${\bf s}$ acting on $\p$ as follows 
\begin{equation}
{\bf s}:\left(\psi(s),\chi_L,\chi_R,v\right)\mapsto \left(s\psi(s),-\lambda\chi_L,\lambda\chi_R, 
{
\Bigl(\begin{array}{cc}\scriptstyle -l/2& \scriptstyle  0\\ \scriptstyle  0& \scriptstyle l/2\end{array}\Bigr)
}
v\right).
\end{equation}
Once we find the orthonormal basis of solutions of $\D^\dagger_t\p=0$ we arrange them as columns of the matrix $\Psi,$ then the orthonormality condition reads $(\Psi,\Psi)={\mathbb I}.$ The instanton connection $\nabla_\mu=\partial_\mu-i A_\mu$ is induced on the kernel of $\D^\dagger_t$ by the connection $D_\mu=\partial_\mu+i{\bf s}a_\mu,$ thus $\nabla_\mu=(\Psi, D_\mu \Psi)$  and the associated  $su(2)$-valued one-form $A=A_0d\tau+A_j dx^j$ is given by
\begin{equation}\label{eq:connection}
A=\left(\Psi,\Bigl(i\frac{\partial}{\partial\tau}-\frac{\bf s}{V}\Bigr)\Psi\right) d\tau+\left(\Psi, \Bigl(i\frac{\partial}{\partial x_j}-\omega_j\frac{\bf s}{V}\Bigr)\Psi\right) dx_j
\end{equation}

\section{The ADHM Limit}
To compare with the ADHM construction we solve $\D_t^\dagger\Psi=0$ at the ends of the interval $s=\pm\frac{l}{2}$ to find
\begin{align}\label{eq:ends}
\psi{\scriptstyle (-\frac{l}{2})}&=-\left(\begin{array}{cc}
B_{10}^\dagger & -b_{10}^\dagger \\
-B_{01} & -b_{01}\end{array}\right) \left(\begin{array}{c} v_{+} \\ v_{-}\end{array}\right),
&
\psi{\scriptstyle(\frac{l}{2})}&=
\left(\begin{array}{cc}
-b_{01}^\dagger & B_{01}^\dagger \\
b_{10} & B_{10}
\end{array}\right) 
\left(\begin{array}{c} v_{+} \\ v_{-}\end{array}\right),
\end{align}
The Nahm equations imply that $\vec{t}$ is constant on $[-l/2,l/2].$  

It is illustrative to consider first the case of a single instanton. In this case $\vec{T}$ is constant on each of the three intervals of ${\cal I}\backslash\{-\lambda,\lambda\}.$ 
Moreover, the values on the left and on the right intervals are equal, thus for some constant vectors $\vec{T}_1$ and $\vec{T}_2$
\begin{equation}\label{eq:piece}
\vec{T}(s)=\begin{cases}
\vec{T}_1 &\text{for}  -l/2<s<-\lambda\ \text{or}\ \lambda>s>l/2,\\
\vec{T}_2 &\text{for}  -\lambda<s<\lambda.
\end{cases}
\end{equation}
Let  $\vec{z}_1=\vec{t}-\vec{T}_1$ and $\vec{z}_2=\vec{t}-\vec{T}_2,$ then the Weyl equation $\D_t^\dagger\Psi=0$  becomes equivalent to 
\begin{multline}
\left[ e^{-\zslash_1(\frac{l}{2}-\lambda_2)} 
\Bigl(\begin{array}{cc} 
\scriptstyle -b_{01}^\dagger & \scriptstyle B_{01}^\dagger \\
\scriptstyle b_{10} & \scriptstyle B_{10}
\end{array}\Bigr)
+ e^{\zslash_2(\lambda_2-\lambda_1)} e^{\zslash_1(\lambda_1+\frac{l}{2})} 
\Bigl(\begin{array}{cc}
\scriptstyle B_{10}^\dagger & \scriptstyle -b_{10}^\dagger \\
\scriptstyle-B_{01} & \scriptstyle -b_{01}
\end{array}\Bigr)
\right]
\left(\begin{array}{c}
v_{+} \\
v_{-}
\end{array}\right)+\\
\label{Eq:ADHM}
+e^{\vsigma\cdot\vz_2(\lambda_2-\lambda_1)}
\Bigl(\begin{array}{c} \scriptstyle J^\dagger_L \\ \scriptstyle I_L\end{array}\Bigr)\chi_L+
\Bigl(\begin{array}{c} \scriptstyle J^\dagger_R \\ \scriptstyle I_R\end{array}\Bigr)\chi_R=0.
\end{multline}
Clearly in the limit of $l\rightarrow 0$ (and since $\lambda<l/2,$ we have $\lambda\rightarrow 0$)  the above expression reduces to the ADHM linear equation.

For the case of instantons of general charge the exponentials in the above equation become path-ordered exponentials involving the corresponding nonabelian data $T$.  Each of these represents parallel transport along an interval. In the $l\rightarrow 0$ limit, however, all of the intervals in the bow diagram contract to a point and the exponential factors all become identities.  Therefore,  for a general case, the equation \eqref{Eq:ADHM} for the kernel of $\D_t$ reduces to the ADHM linear equation.

\section{Proof of Self-duality}
The core of this proof is close to the original argument of Nahm \cite{Nahm:1979yw}, but requires some adjustments.  We shall need the following relations
\footnote{These follow from Eq.(\ref{Eq:b-relations}) and $b_\pm b_\pm^\dagger=t\pm\tslash.$ Namely,  (\ref{Eq:b-relations}) implies $4itVa b_-=2b_-(db_-^\dagger) b_- -b_-d(b_-^\dagger b_-)=2(d(b_-b_-^\dagger))b_--2(db_-)(b_-^\dagger b_-)-b_-d(b_-^\dagger b_-)=2(dt-d\tslash)b_--4tdb_--2dt b_-=-4tdb_--2d\tslash\, b_-,$ thus $2t(db_-+ila b_-)=-(d\tslash+ia)b_-.$}
 \begin{align}
 (d+i l a)b_-&=-\frac{1}{2t}\big(d\tslash+ia\big) b_-& \text{and}&& (d-il a)b_+&=\frac{1}{2t}\big(d\tslash+ia\big)b_+.
 \end{align}
We also use the fact that $\D^\dagger_t\D_t=1\otimes\Delta,$ with $\Delta$ positive definite (except for some $(\tau, \vec{t})$ corresponding to a finite number of isolated points on the Taub-NUT).  Thus, it has a well defined inverse $G=\big(\D^\dagger_t\D_t\big)^{-1},$ given by the Green's function of $\Delta,$ that commutes with the quaternions and the $\sigma$-matrices.

As expressed by Eq.~(\ref{eq:connection}), the connection induced by $D_\mu$ on the kernel of $\D^\dagger$ is the instanton connection $A_\mu,$ therefore the covariant differential is  $dt^\mu \nabla_\mu\equiv dt^\mu(\partial_\mu-i A_\mu)=Pdt^\mu D_\mu P=P(d+i s a)P=P\big(d+i s\frac{d\tau+\omega}{2V}\big)P.$
So the connection is $A_\mu=i(\Psi, D_\mu \Psi),$ then
\begin{eqnarray}
\partial_{[\mu} A_{\nu]}&=&i \big(D_{[\mu}\Psi, D_{\nu]}\Psi \big)-(\Psi, {\bf s}\Psi)\partial_{[\mu} a_{\nu]}\\
{}[A_\mu, A_\nu] &=& \big(D_{[\mu}\Psi, \Psi \big) \big(\Psi, D_{\nu]} \big).
\end{eqnarray}
It follows that the curvature is 
\begin{equation}
\label{curvature}
F_{\mu\nu}=i[\nabla_\mu, \nabla_\nu]= \big(D_{[\mu}\Psi, (1-P) D_{\nu]}\Psi)-(\Psi,{\bf s}\Psi \big)\partial_{[\mu}a_{\nu]},
\end{equation}
where $1-P\equiv 1-\Psi\Psi^\dagger=\D_t G \D_t^\dagger.$  The second term in the curvature expression is self-dual since $da$ is, while for the first term we have
\begin{equation}\label{relation}
(D_{\mu}\Psi, (1-P) D_{\nu}\Psi)=\big([\D_t^\dagger, D_\mu]\Psi, G [\D_t^\dagger, D_\nu]\Psi\big).
\end{equation}

Since
\begin{equation}
\begin{split}
\D_t^\dagger&=\bigg(\frac{d}{ds}+i T_0+\Tslash-\tslash\bigg)\oplus\delta(s-\lambda_\alpha) Q_\alpha\\
&\quad \oplus\Bigl(\delta(s+l/2)(B_-, -b_+)+\delta(s-l/2)(-b_-, B_+)  \Bigr).
\end{split}
\end{equation}
The commutator $dt^\nu [\D_t^\dagger, D_\nu]$ is given by
\begin{equation}\label{Eq:comm}
\begin{split}
[\D_t^\dagger, d+i \mathbf{s} a]&=\big(d\tslash+ia\big)\oplus 0\oplus\Big(\delta(s-l/2)(d+i l a)b_-,\, \delta(s+l/2)(d-i l a)b_+\Big)\\
  &=\big(d\tslash+ia\big)\bigg(1\oplus 0\oplus\Big(-\frac{1}{2t}\delta(s-l/2) b_-, \frac{1}{2t}\delta(s+l/2) b_+\Big)\bigg).
\end{split}
\end{equation}
As the Green's function $G=(\D^\dagger \D)^{-1}$ is scalar, that is, it commutes with the $\sigma$-matrices, and $\big(d\tslash-ia\big)\wedge\big(d\tslash+ia\big)= \big(\frac{1}{2}d\xslash+ia \big)\wedge \big(\frac{1}{2}d\xslash-ia \big)$ is self-dual according to Eq.(\ref{Eq:SDForms}), the curvature two-form $F=F_{\mu\nu}dx^\mu dx^\nu$ is self-dual as well due to Eqs.(\ref{curvature},\ref{relation},\ref{Eq:comm}).

\section{Instantons for the $U(n)$ Gauge Group}
\begin{figure}[htbp]
\begin{center}
\includegraphics[width=0.5\textwidth]{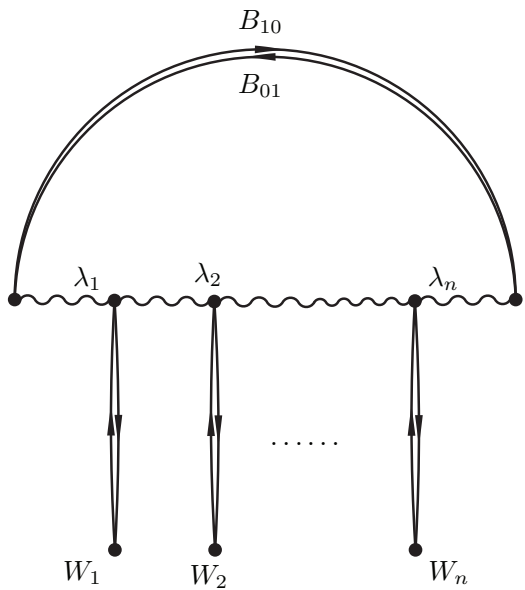}
\caption{Bow diagram for $U(n)$ Instanton on Taub-NUT.}
\label{fig:UnInst}
\end{center}
\end{figure}

Generalizing our construction to instantons with the gauge group $U(n)$ is fairly straightforward. 
The corresponding bow diagram is given in Figure \ref{fig:UnInst}.  The positions of the marked points $\lambda_1,\ldots, \lambda_n$ partitioning the interval $[-l/2, l/2]$ are given by the asymptotic of the eigenvalues of the instanton connection monodromy around the Taub-NUT circle. All of our previous discussion including the proof of the self-duality and the ADHM limit remains valid.

\section{The Geometric Meaning of the Nahm Transform for Curved  Manifolds}
The conventional Nahm transform \cite{DK} of some self-dual configuration (or of a dimensional reduction of a self-dual configuration) on a flat manifold $M={\mathbb R}^4/\Lambda$ results in some data on a dual space $N$ of flat connections on $M$.  The kernel of the Nahm transform is the Poincar\'e bundle ${\mathfrak P}\rightarrow M\times N.$  Let us denote the two projections of the product $M\times N$ on $M$ and $N$ by $p_M$ and $p_N$ respectively, so that we have the following diagram
\begin{equation}
\xymatrix{
 &{\mathfrak P}\ar[d]& \\
 &\ar[dl]_{p_M}M\times N\ar[dr]^{p_N}& \\
M& &N.
}
\end{equation}
Then for an instanton bundle ${\cal E}\rightarrow M$ its Nahm transform is $p_{N*}\big({\mathfrak P}\otimes p^*_M {\cal E}\big).$  
Thus the Poincar\'e bundle ${\mathfrak P}$ plays the role of the kernel of this transform. 

For example, for the case of a caloron, flat connections on ${\mathbb R}^3\times S^1$ have the form $\eta=s dt_0,$ where $t_0$ is the coordinate parameterizing the $S^1$ factor.  The space of such connections forms the dual circle $\hat{S}^1$ parameterized by $s.$  The Poincar\'e bundle over the product $({\mathbb R}^3\times S^1)\times\hat{S}^1$ has a natural connection $\eta$ with curvature $\F=ds\wedge dt_0,$ and it can be trivilalized on either one of the two base components, making both pushforward operations $p_{N*}$ and $p_{M*}$ simple and well defined.
\begin{equation}\label{NahmMonDiag}
\xymatrix{
 &{\mathfrak P}\ar[d]& &\F=d\eta=ds\wedge dt_0 \\
 &\ar[dl]_{p_M}\big({\mathbb R}^3\times S^1\big)\times \hat{S}^1\ar[dr]^{p_N}& &\\
\stackrel{\vec{t},\ t_0}{{\mathbb R}^3\times S^1}& \eta=s dt_0 &\stackrel{s}{\hat{S}^1}.& 
}
\end{equation}

For a curved manifold $M$ without flat connections, such as the Taub-NUT space, a generalization of the Nahm transform is less straightforward.  In order to have a version of the Nahm transform in the diagram \eqref{NahmDiagram} below, one has to answer two questions: 1) What is the correct choice of the `dual' manifold $N$? and 2) What is the kernel ${\mathfrak{M}}$ generalizing the Poincar\'e bundle?
\begin{equation}\label{NahmDiagram}
\xymatrix{
 &{\mathfrak{M}}\ar[d]& \\
 &\ar[dl]_{p_M}(\mbox{\rm Taub-NUT})\times {\cal I} \ar[dr]^{p_N}& \\
\mbox{\rm Taub-NUT}& \eta= s \frac{d\tau+\omega}{2V} &{\cal I}.
}
\end{equation}
We propose that for the Taub-NUT space the appropriate choice of $N$ is the space ${\cal I}$ of self-dual Abelian connections on the Taub-NUT (or rather, in order to have a hyperk\"ahler space, the direct product of ${\cal I}$ and ${\mathbb R}^3$).  In order to answer the second question, we digress to discuss a generalization of instantons in four dimensions to instantons on higher-dimensional spaces.

Instantons on higher dimensional hyperk\"ahler manifolds were defined in \cite{MCS} in the following manner.  Consider the operator $\aleph=I\otimes I+J\otimes J+K\otimes K$ acting on two-forms.  Due to the defining quaternionic identities it satisfies
\begin{equation}
\aleph^2=2\aleph+3,
\end{equation}
and can have eigenvalues $3$ or $-1.$
On a four-dimensional hyperk\"ahler manifold this operator is related to the Hodge star operation by $*=\frac{1}{2}(\aleph-1),$  
thus on a general hyperk\"ahler manifold the equations 
\begin{align}
\aleph \F&=3 \F &&\text{and} & \aleph \F&=-\F
\end{align}
respectively generalize the self-duality and anti-self-duality conditions to higher dimensions.

Before we proceed, let us observe that the complex structures $I, J,$ and $K$  act on the vierbein $e^{\hat{\mu}}$ of the Taub-NUT
\begin{align}
e^{\hat{0}}&=\frac{1}{2\sqrt{V}}(d\tau+\omega),& e^{\hat{j}}&=\frac{1}{2}\sqrt{V} dx^{\hat{j}},
\end{align}
by acting with the left multiplication on the quaternionic combination $e^{\hat{0}}+I e^{\hat{1}}+J e^{\hat{2}}+K e^{\hat{3}}.$

Just as the four-dimensional self-duality equations become Bogomolny equations under the reduction to three dimensions, we reduce an eight-dimensional self-duality condition $3 \F=\aleph \F$ to five dimensions producing the following system of equations on ${\cal I}\times$Taub-NUT:
\begin{align}
3 \F_{\hat{0}s}&=\hat{D}_1\Phi_1+\hat{D}_2\Phi_2+\hat{D}_3\Phi_3,\nonumber\\
3 \F_{\hat{1}s}&=-\hat{D}_0\Phi_1-\hat{D}_3\Phi_2+\hat{D}_2\Phi_3,\nonumber\\
\label{Eq:Monotone}
3 \F_{\hat{2}s}&=\hat{D}_3\Phi_1-\hat{D}_0\Phi_2-\hat{D}_1\Phi_3,\\
3 \F_{\hat{3}s}&=-\hat{D}_2\Phi_1+\hat{D}_1\Phi_2-\hat{D}_0\Phi_3,\nonumber\\
2\F_{\hat{\mu}\hat{\nu}}&=\epsilon_{\hat{\mu}\hat{\nu}\hat{\rho}\hat{\sigma}} \F_{\hat{\rho}\hat{\sigma}}.\nonumber
\end{align}
Here $\Phi_1, \Phi_2,\Phi_3$ are the components of the eight-dimensional connection in the reduced three directions of ${\mathbb R}^3.$  We used the curvature vierbein components $\F=\F_{s\hat{\rho}} ds\wedge e^{\hat{\rho}}+\F_{\hat{\mu}\hat{\nu}} e^{\hat{\mu}}\wedge e^{\hat{\nu}}$  and $\hat{D}_0=2\sqrt{V}D_0$ and $\hat{D}_j=\frac{2}{\sqrt{V}}D_j-4\omega_j D_0,$ which appear in the covariant differential decomposition  $D=d\tau D_0+dx^j D_j=e_{\hat{0}} \hat{D}_{\hat{0}}+e^{\hat{j}}\hat{D}_{\hat{j}}.$

Since Eqs.(\ref{Eq:Monotone}) emerge via a dimensional reduction of higher-dimensional self-duality equations, one might call an object satisfying these equations an {\em Instapole} or a {\em Monotone}\footnote{We hope someone will come up with a more poetic  name for it.}.

In our case $\F=d\eta=d\left(s\frac{d\tau+\omega}{2V}\right)=ds\wedge a+s da=\frac{1}{\sqrt{V}}ds\wedge e^0+s da$ as dictated by the relation (\ref{NahmMonDiag}) between $\cal I$ and the Taub-NUT.  We observe that 
$\Phi_1=t_1=-\frac{1}{2} x_1, \Phi_2=t_2=-\frac{1}{2} x_2, \Phi_3=t_3=-\frac{1}{2} x_3$ augment $\F$ to produce a solution to the system of Eq.~(\ref{Eq:Monotone}).
This is exactly the solution defining the object generalizing the Poincar\'e bundle in the diagram (\ref{NahmDiagram}) that leads to the twisting that we used in Eq.(\ref{Twist}).  It plays the role of the kernel in this generalization of the Nahm transform. 

\section{Example of One Instanton}
Let us now focus on a single $SU(2)$ instanton on the Taub-NUT, i.e. a self-dual curvature configuration with $k_0=1$ and $m=0.$
For a single instanton the $T$'s in the Nahm data are Abelian and the Nahm equations are solved by 
\begin{equation}
\vec{T}(s)=\begin{cases}
\vec{T}_1 &\text{for  $-l/2<s<-\lambda$ or $\lambda>s>l/2,$}\\
\vec{T}_2 &\text{for  $-\lambda<s<\lambda.$}
\end{cases}.
\end{equation}
We interpret $\vec{x}=-2\vec{T}_1$ and $\vec{x}=-2\vec{T}_2$ as the locations of the instanton constituents. Let $\vec{z}_{1}=\vec{t}-\vec{T}_{1}$ and  $\vec{z}_{2}=\vec{t}-\vec{T}_{2}$ denote the position relative to the two constituents and let $\vec{y}=\vec{T}_2-\vec{T}_1=\vec{z}_1-\vec{z}_2$ be the displacement between them.  The $\tau$ coordinate of the instanton position is proportional to $T_0.$  Since the Taub-NUT metric is invariant with respect to shifts of $\tau,$ without loss of generality $T_0$ can be put to zero.
We also gauge away $t_0$ in favor of the phase of $b_\pm.$  
Let the two-component spinors $Q_+$ and $Q_-$ be such that $Q_\pm Q_\pm^\dagger=y\pm\yslash.$
Using the component expressions for the spinors we introduced earlier 
\begin{equation}
b_-=
\left(
\begin{array}{r}
b^\dagger_{01}\\
-b_{10}  
\end{array}
\right),\ 
b_+=
\left(
\begin{array}{r}

b^\dagger_{10}\\
b_{01}  
\end{array}
\right),\ 
B_-=
\left(
\begin{array}{r}
B^\dagger_{10}\\
-B_{01}  
\end{array}
\right),\ 
B_+=
\left(
\begin{array}{r}
B^\dagger_{01}\\
B_{10}  
\end{array}
\right),
\end{equation}
and, since in this case all of the components are simply complex numbers, it is straightforward to verify that
\begin{equation}
b^\dagger_-B_-=B^\dagger_+b_+=e^{ i\tau/2}{\cal P},\ 
b^\dagger_+ B_+=B^\dagger_- b_-=e^{-i\tau/2}{\cal P},
\end{equation}
where
\begin{equation}\label{Eq:P}
{\cal P}=\sqrt{(T_1+t)^2-z_1^2}.
\end{equation}
 The moment maps at $s=\pm l/2$  imply that 
\begin{equation}
b_\pm b^\dagger_\pm=|\vec{t}\,|\pm\tslash\ , \ 
B_\pm B^\dagger_\pm=|\vec{T}_1|\pm\Tslash_1,\ 
\end{equation}
and the vanishing of the moment maps at $s=\pm\lambda$ implies $Q_R=Q_+$ and $Q_L=Q_-.$

\subsection{Solving the Weyl Equation}\label{Sec:Solution}
On each interval $\D_t=-\partial_s+\Tslash-\tslash=-\partial_s-\zslash$ and $\D^\dagger_t=\partial_s+\Tslash-\tslash=\partial_s-\zslash,$ with $\zslash=\zslash_1$ or $\zslash_2$ in accordance with Eq.~(\ref{eq:piece}).
It follows therefore that $\psi(s)$ has the form
\begin{equation}\label{eq:psi}
\psi(s)=\begin{cases}
e^{\zslash_1(s+l/2)}\psi_L  &\text{for $-l/2<s<-\lambda,$}\\
e^{\zslash_2 s} \Pi  &\text{for $-\lambda<s<\lambda,$}\\
e^{\zslash_1(s-l/2)}\psi_R  &\text{for $\lambda<s<l/2,$} 
\end{cases}
\end{equation}
for some constant $\psi_L, \psi_R,$ and $\Pi.$   As we shall soon verify, the kernel of $\D_t^\dagger$ is two-dimensional, so from now on we shall understand $\psi(s), \chi_{L}, \chi_R$ and $v$ to be two-column matrices, so that their first columns deliver one of the solutions and their second columns deliver the second, linearly independent, solution of $\D_t^\dagger\p=0.$

Let $A_L=(B_-, -b_+)$ and $A_R=(-b_-, B_+)$ 
then the $\D_t^\dagger\Psi=0$ conditions read
\begin{align}
\label{Lend}
\psi_R-A_R v&=0&&\text{at $s=\frac{l}{2}$}\\
\label{MidL}
e^{-\zslash_1 (l/2-\lambda)}\psi_R-e^{\zslash_2\lambda}\Pi+Q_R\chi_R&=0&&\text{at $s=\lambda,$}\\
\label{MidR}
e^{-\zslash_2\lambda}\Pi-e^{\zslash_1 (l/2-\lambda)}\psi_L+Q_L\chi_L&=0&&\text{at $s=-\lambda,$}\\
\label{Rend}
\psi_L+A_L v&=0&&\text{at $s=-\frac{l}{2}.$}
\end{align}
It is useful to note the following relations
\begin{alignat}{3}
\nonumber
A_L A^\dagger_L&=T_1+t+\zslash_1, & A^\dagger_R A_L&=A_L A^\dagger_R=-e^{i\tau/2}{\cal P}, &  A_L A_R^{-1}&=-\frac{e^{i\tau/2}}{{\cal P}}(T_1+t+\zslash_1),\\
\nonumber
A_R A^\dagger_R&=T_1+t-\zslash_1, & A^\dagger_L A_R&=A_R A^\dagger_L=-e^{-i\tau/2}{\cal P}, & A_R A_L^{-1}&=-\frac{e^{-i\tau/2}}{{\cal P}}(T_1+t-\zslash_1),
\end{alignat}
and define $\mu_+$ and $\mu_-$ to be such that $\mu^2_+=A_L A_L^\dagger$ and $\mu^2_-=A_R A_R^\dagger$ namely
\begin{equation}\label{eq:mu}
\mu_\pm=\sqrt{\frac{T_1+t+{\cal P}}{2}}\pm\sqrt{\frac{T_1+t-{\cal P}}{2}}\frac{\zslash_1}{z_1},
\end{equation}
then $\mu_+\mu_-={\cal P}.$

We choose 
\begin{equation}
\label{eq:v}
v=-e^{i\tau/4} A_L^\dagger\frac{\mu_-}{{\cal P}}=e^{-i\tau/4}A_R^\dagger\frac{\mu_+}{{\cal P}},
\end{equation} 
so that now $\psi_L=e^{i\tau/4}\mu_+,\ \psi_R=e^{-i\tau/4}\mu_-.$ From the matching conditions Eqs.(\ref{MidL}) and (\ref{MidR}) at $s=\pm\lambda$ it follows that
\begin{equation}\label{eq:Pi}
\Pi=\frac{1}{2g}\left(e^{-i\tau/4} e^{\lambda\zslash_2}(y-\yslash) e^{-(l/2-\lambda)\zslash_1}\mu_-+
e^{i\tau/4} e^{-\lambda\zslash_2}(y+\yslash) e^{(l/2-\lambda)\zslash_1}\mu_+\right), 
\end{equation}
where the function $g$ is given by 
\begin{equation}\label{Eq:g}
g=y \cosh 2z_2\lambda-\frac{\vec{z}_2\cdot\vec{y}}{z_2}\sinh 2z_2\lambda=\frac{1}{2}\left(e^{2\zslash_2\lambda}Q_-Q^\dagger_-+Q_+Q^\dagger_+e^{-2\zslash_2\lambda}\right),
\end{equation}
and that
\begin{equation}\label{eq:chi}
\left(\begin{array}{c} \chi_R\\ \chi_L\end{array}\right)=\left(\begin{array}{c} Q_+^\dagger e^{-\lambda\zslash_2}\\ Q_-^\dagger e^{\lambda\zslash_2}\end{array}\right) \Upsilon,
\end{equation}
with 
\begin{equation}\label{Eq:Upsilon}
\Upsilon=\frac{e^{i\tau/4}e^{\lambda\zslash_2}e^{(\frac{l}{2}-\lambda)\zslash_1}\mu_+-e^{-i\tau/4}e^{-\lambda\zslash_2}e^{-(\frac{l}{2}-\lambda)\zslash_1}\mu_{-}}{2g}.
\end{equation}

\subsection{Normalization}
Let us now check the orthogonality and the normalization of the solution  $\Psi$ delivered by Eqs.(\ref{eq:psi}, \ref{eq:v}, \ref{eq:chi}).  To simplify our notation let us introduce  $\alpha=\frac{1}{4z_1}\ln\frac{T_1+t+z_1}{T_1+t-z_1},$ so that   $\mu_-^2=T_1+t-\zslash_1={\cal P}e^{-2\alpha\zslash_1},$ and in particular that $\sinh2\alpha z_1=z_1/{\cal P}$ and $\cosh2\alpha z_1=(T_1+t)/{\cal P}.$ 
 Introduce $\Delta=\frac{l}{2}-\lambda+\alpha$ and let
\begin{eqnarray}
\label{Eq:Cosh}
c_1&=&\cosh 2 \Delta z_1=\frac{(T_1+t)\cosh(l-2\lambda)z_1+z_1\sinh(l-2\lambda)z_1}{{\cal P}}, \\
\label{Eq:Sinh}
s_1&=&\sinh 2 \Delta z_1=\frac{z_1\cosh(l-2\lambda)z_1+(T_1+t)\sinh(l-2\lambda)z_1}{{\cal P}}, \\
\label{Eq:CoSinh}
c_2&=&\cosh 2\lambda z_2,\ \ s_2=\sinh 2\lambda z_2.
\end{eqnarray}
In these terms $g=y c_2+\frac{y^2+z_2^2-z_1^2}{2z_2} s_2.$ Then  we find that $(\Psi,\Psi)=N^2 {\mathbb I}_{2\times 2}$with the normalization factor 
\begin{equation}
\label{Eq:Norm}
N^2=(\Psi, \Psi)=\frac{{\cal P}}{g}\Big( c_1 c_2+\frac{y}{z_1} s_1 c_2+\frac{y}{z_2} c_1 s_2+\frac{z_1^2+z_2^2+y^2}{2z_1 z_2} s_1 s_2-\cos \frac{\tau}{2}\Big).
\end{equation}

Let us also observe that both $\Pi$ and $\Upsilon$ appearing in the solution are scalar multiples of unitary matrices, since 
\begin{align}
\Pi^\dagger\Pi&=\frac{{\cal P}}{g}\left(y c_1+\frac{\vec{y}\cdot\vec{z}_1}{z_1}s_1\right),& 
\Upsilon^\dagger\Upsilon&=\frac{{\cal P}}{2 g^2}\left(c_1 c_2+s_1 s_2\frac{\vec{z}_2\cdot\vec{z}_1}{z_1 z_2}-\cos\frac{\tau}{2}\right).
\end{align}

\subsection{Connection}
We rewrite Eq.(\ref{eq:connection}) as
$A=A^{(0)} d\tau+A^{(3)}-\Phi\frac{1}{2V}(d\tau+\vec{\omega}\cdot d\vec{x}),$ where
\begin{align}
\Phi&=(\Psi_N,{\bf s}\Psi_N),&  A^{(0)}&=i\left(\Psi_N,\frac{\partial}{\partial\tau}\Psi_N\right),& A^{(3)}&=i\left(\Psi_N,\frac{\partial}{\partial x_j}\Psi_N\right) d x^j,
\end{align}
for an orthonormalized solution $\Psi_N$ satisfying $(\Psi_N, \Psi_N)={\mathbb I}_{2\times 2}.$  Here we observe that for any $\Psi=N \Psi_N,$ with $N$ any nowhere vanishing scalar function, we have  
\begin{multline}
(\Psi_N,\frac{\partial}{\partial x^\mu}\Psi_N)=\frac{1}{2}\left((\Psi_N,\frac{\partial}{\partial x^\mu}\Psi_N)-(\frac{\partial}{\partial x^\mu}\Psi_N,\Psi_N)\right)\\
=\frac{1}{2 N^2}\left((\Psi,\frac{\partial}{\partial x^\mu}\Psi)-(\frac{\partial}{\partial x^\mu}\Psi,\Psi)\right).
\end{multline}  
Thus for the solution of Section \ref{Sec:Solution} which satisfies $(\Psi, \Psi)=N^2 {\mathbb I}_{2\times 2}$ we have
\begin{align}
\Phi&=\frac{1}{N^2}(\Psi,{\bf s}\Psi),\\  
A^{(0)}&=\frac{i}{N^2}(\Psi, \frac{\partial}{\partial\tau}\Psi)=\frac{i}{2N^2}\left((\Psi, \frac{\partial}{\partial\tau}\Psi)-( \frac{\partial}{\partial\tau}\Psi, \Psi)\right),\\
A^{(3)}&=\frac{i}{N^2}(\Psi, d \Psi)=\frac{i}{2N^2}\Big(\Psi, (d-\overleftarrow{d}) \Psi\Big)\equiv\frac{i}{2N^2}\big((\Psi, d \Psi)-( d \Psi, \Psi)\big),
\end{align}
where we introduced the three-dimensional differential $d=dx^j\frac{\partial}{\partial x^j}=dt^j\frac{\partial}{\partial t^j}.$

Given our solution for $\Psi$ of Eqs.(\ref{eq:psi}, \ref{eq:v}, \ref{eq:chi}) one can apply the above formulas, performing some elementary integrals over $s$. 
A straightforward if tedious calculation gives
\begin{multline}
N^2 A^{(0)}=
\frac{1}{4}\left(2t-c_1{\cal P}\right)\frac{\zslash_1}{z_1^2}+\frac{1}{2{\cal P}}\left(\Tslash_1-\frac{\vec{T}_1\cdot\vec{z}_1}{z_1}\frac{\zslash_1}{z_1}\right)\\
\label{Eq:A0}
\quad +\frac{i}{2}\frac{s_2}{z_2}\Pi^\dagger(\partial_\tau-\pl_\tau)\Pi+ig\Upsilon^\dagger(\partial_\tau-\pl_\tau)\Upsilon,
\end{multline}
\begin{multline}
N^2\Phi=\left(1+2 l t-\Big(2\lambda c_1+\frac{s_1}{z_1}\Big){\cal P}\right)\frac{\zslash_1}{2z_1^2}
+\frac{l}{{\cal P}}\left(\Tslash_1-\frac{\Tslash_1\cdot\zslash_1}{ z_1}\frac{\zslash_1}{z_1}\right)\\
+\left(2\lambda c_2-\frac{s_2}{z_2}\right)\Pi^\dagger\frac{\zslash_2}{2z_2^2}\Pi+2\lambda\Upsilon^\dagger\left(\big\{(c_2-1)\vec{y}\cdot\vec{z}_2-s_2 y z_2\big\}\frac{\zslash_2}{z_2^2}+\yslash\right)\Upsilon,
 \label{Eq:Phi}
\end{multline}
\begin{eqnarray}
N^2 A^{(3)}&=&\frac{i}{2}\left\{
\frac{z_1}{{\cal P}^2}\left[\Tslash_1, d\frac{\zslash_1}{z_1}\right]+\frac{1}{{\cal P}^3}\left(\frac{T_1+t}{z_1}\frac{\vec{z}_1\cdot d\vec{t}}{z_1}-\frac{\vec{t}\cdot d\vec{t}}{t}\right)[\Tslash_1, \zslash_1]\right.\nonumber\\
&&\qquad -\left(1+{\cal P}\left(l-2 \lambda-\frac{s_1}{z_1}\right)-
2 \frac{T_1(T_1+t-{\cal P})}{{\cal P}^2}
\right)\frac{[\zslash_1, d\tslash]}{2 z_1^2}\nonumber\\
&&\qquad +\Pi^\dagger\left(\frac{s_2}{z_2}d-\overleftarrow{d}\frac{s_2}{z_2}-\Big(2\lambda-\frac{s_2}{z_2}\Big)\frac{[\zslash_2, d\tslash]}{2 z_2^2}\right)\Pi\nonumber\\
&&\qquad +\Upsilon^\dagger\left(2g d-\overleftarrow{d} 2g-2\lambda\frac{\vec{z}_2\cdot d\vec{t}}{z_2^2}[\yslash, \zslash_2]\right.\nonumber\\
\label{Eq:A3}
&&\qquad \phantom{+\Upsilon^\dagger\left(\right.}\ \left.\left. 
-s_2\left[\yslash, d\frac{\zslash_2}{z_2}\right]+(c_2-1)\frac{y}{z_2^2}[\zslash_2, d\tslash]\right)\Upsilon
\right\},
\end{eqnarray}
where the functions ${\cal P}$ and $g$ are defined in Eqs.~\eqref{Eq:P}  and \eqref{Eq:g}, the hyperbolic functions $c_1, s_1, c_2$ and $s_2$ are in Eqs.~(\ref{Eq:P}, \ref{Eq:Sinh}) and \eqref{Eq:CoSinh}, and  $\Pi$ and $\Upsilon$ are given in Eqs.~\eqref{eq:Pi} and \eqref{Eq:Upsilon}. The normalization factor $N^2$ is read from Eq.~\eqref{Eq:Norm}.

\section{Conclusions}
We discussed topological charges of an instanton configuration on the Taub-NUT space with the maximal symmetry breaking by the monodromy at infinity.  These are given by integer monopole charges and an integer instanton number.  Solutions with vanishing instanton number correspond to singular monopoles \cite{Kronheimer:1985}.  In their three-dimensional interpretation these have infinite energy, while as configurations on the Taub-NUT space they are smooth and  have finite action.  Thus one can regard the Taub-NUT background as a regularization.  A simplest solution with zero  instanton number was constructed in \cite{Cherkis:2007qa} and its physical properties were explored in \cite{Cherkis:2007jm}.    

In this manuscript we focussed on the case with vanishing monopole charges.  We presented the ADHM-Nahm data for this case.  These data are conveniently encoded in a bow diagram, such as in Figure \ref{fig:Instanton} or Figure \ref{fig:UnInst}.  We used the bow diagram description earlier in \cite{Cherkis:2008ip} to study the moduli spaces of instantons on the Taub-NUT.  Here we give the details of the Nahm transform leading to the explicit instanton connection.

As an example illustrating this construction we find a single $SU(2)$ instanton on the Taub-NUT space in Eqs.(\ref{Eq:A0}, \ref{Eq:Phi}, \ref{Eq:A3}).

The bow diagram formalism we presented is not limited to the case of the Taub-NUT background.  Rather, we chose to limit the scope of this paper to this case to simplify our presentation.  In the forthcoming paper \cite{Cherkis:2010bn} we will give the bow-diagrammatic description of instantons with arbitrary charges on a general  ALF spaces of either $A_k$- or $D_k$-type.

\section*{Acknowledgments}
It is our pleasure to thank Tamas Hausel and  Juan Maldacena for a number of useful conversations. We are grateful to Christopher Blair for careful reading of the manuscript and for identifying a number of misprints in its original version. This work is supported by Science Foundation Ireland Grant No. 06/RFP/MAT050 and by the European Commission FP6 program MRTN-CT-2004-005104. 
\pagebreak

\section*{Appendix}
\subsection*{Metric Conventions and Moment Maps}
The Nahm data on an interval of length $l$ can be organized into a quaternion 
$t=t_0 e_0+\vec{t}\cdot\vec{e}$ with the metric and the symplectic forms
\begin{equation}\nonumber
ds^2=g(\cdot,\cdot)=l\frac{1}{2}{\rm tr} dt dt^\dagger,\ 
\omega_j(A,B):=g(A,e_j B)=-\frac{l}{4}{\rm tr} e_j(A B^\dagger-B A^\dagger),
\end{equation}
\begin{equation}
\omega_j=g(\cdot,e_j\cdot)=-\frac{l}{4}{\rm tr}\big(dt\wedge dt^\dagger e_j\big).
\end{equation}
With respect to $t\mapsto t+\epsilon$ the moment maps are
$\mu_j=-\frac{l}{4}{\rm tr} (t^\dagger-t)e_j=l t_j.$

For $q=(b_-,b_+)$ the metric is $ds^2=\frac{1}{2}{\rm tr} dq dq^\dagger=db_-^\dagger db_-=db_+^\dagger db_+.$ The moment map with respect to $q\mapsto q\, e^{\epsilon e_3}$ is $\mu_j=-\frac{1}{4}{\rm tr} q e_3 q^\dagger e_j.$

Coordinates on Taub-NUT are either $b_+, b_-$ or $\vec{t}, 2 l t_0=\tau\sim\tau+4\pi.$
The instanton moduli are $\vec{T}_1, \theta_1=(2l-4\lambda)T^L_0$ (or $B_+, B_-$) and $\vec{T}_2, \theta_2=4\lambda T^M_0.$
The relative coordinates are $\vec{z}_1=\vec{t}-\vec{T}_1, \vec{z}_2=\vec{t}-\vec{T}_2,$ and the relative position is $\vec{y}=\vec{T}_2-\vec{T}_1=\vec{z}_1-\vec{z}_2.$

We also collect the bifundamental data as
\begin{equation}
b_-=\left(\begin{array}{c}b_{01}^\dagger\\-b_{10}\end{array}\right),\ 
b_+=\left(\begin{array}{c}b_{10}^\dagger\\ b_{01}\end{array}\right),\ 
B_-=\left(\begin{array}{c}B_{10}^\dagger\\-B_{01}\end{array}\right),\ 
B_+=\left(\begin{array}{c}B_{01}^\dagger\\ B_{10}\end{array}\right),
\end{equation}
and the fundamental data as
\begin{equation}
Q_-=\left(\begin{array}{c}J_L^\dagger\\ I_L\end{array}\right),\ 
Q_+=\left(\begin{array}{c}J_R^\dagger\\ I_R\end{array}\right).
\end{equation}
\subsection*{Vanishing Moment Map Conditions}
For the Taub-NUT
\begin{equation}
\frac{d}{ds}\tslash+\V\Bigg\{\delta\bigg(s+\frac{l}{2}\bigg)b_-b_-^\dagger+\delta\bigg(s-\frac{l}{2}\bigg)b_+b_+^\dagger\Bigg\}=0,
\end{equation}
and for the instanton Bow Data
\begin{multline}
\Big[\frac{d}{ds} + i T_0,\Tslash\Big]+\V\Bigg\{\Tslash\Tslash+\delta\bigg(s+\frac{l}{2}\bigg)B_-B_-^\dagger+\delta\bigg(s-\frac{l}{2}\bigg)B_+B_+^\dagger \\
+\delta(s+\lambda)Q_-Q_-^\dagger+\delta(s-\lambda)Q_+Q_+^\dagger\Bigg\}=0.
\end{multline}
These imply that $\vec{T}$ is constant on each interval and equals $\vec{T}_1$ for $|s|>\lambda$ and $\vec{T}_2$ for $|s|<\lambda.$  The conditions at $s=l/2, -l/2, \lambda, -\lambda$ are respectively
\begin{equation}
T_1+\Tslash_1=B_+B_+^\dagger,\ T_1-\Tslash_1=B_-B_-^\dagger,\ 
y+\yslash=Q_+Q_+^\dagger,\ y-\yslash=Q_-Q_-^\dagger.
\end{equation}

\subsection*{The Weyl Equation}
\begin{equation}
\Big(\frac{d}{ds}-\zslash_{1,2}\Big)\psi(s)=0,
\end{equation}
\begin{align}
\psi(\lambda+)-\psi(\lambda-)&=-Q_+\chi_R,&
\psi(l/2)=(-b_-, B_+) v,\\ 
\psi(-\lambda+)-\psi(-\lambda-)&=-Q_-\chi_L,& 
\psi(-l/2)=-(B_-, -b_+) v.
\end{align}

\subsection*{Solution of the Weyl Equation}
\begin{align}
v&=\frac{1}{P}\left(\begin{array}{r}
-e^{i\frac{\tau}{4}}  B_-^\dagger\mu_-   \\
e^{-i\frac{\tau}{4}} B_+^\dagger\mu_+\end{array}\right), & 
\left(\begin{array}{c}
\chi_R\\ \chi_L
\end{array}\right)&=\left(\begin{array}{l}
Q_+^\dagger e^{-\lambda\zslash_2}\\
Q_-^\dagger\, e^{\lambda\zslash_2}
\end{array}\right) \Upsilon,
\end{align}
\begin{equation}
\psi(s)=\begin{cases}
e^{-i\frac{\tau}{4}}e^{(s-\frac{l}{2})\zslash_1}\mu_- &\text{for  $\lambda<s<l/2,$}\\
e^{s\zslash_2}\Pi &\text{for $-\lambda<s<\lambda$,}\\
e^{i\frac{\tau}{4}}e^{(s+\frac{l}{2})\zslash_1}\mu_+ &\text{for $-l/2<s<-\lambda.$}
\end{cases}
\end{equation}
Here
\begin{align}
2g&=2\big(y\cosh2\lambda z_2-\frac{\vec{y}\cdot\vec{z}_2}{z_2}\sinh2\lambda z_2\big),& {\cal P}&=\sqrt{(T_1+t)^2-z_1^2},
\end{align}
\begin{equation}
\mu_\pm=\sqrt{\frac{T_1+t+{\cal P}}{2}}\pm\sqrt{\frac{T_1+t-{\cal P}}{2}}\frac{\zslash_1}{z_1},
\end{equation}
and
\begin{align}
\Upsilon&=\frac{1}{2g}\left\{
e^{i\frac{\tau}{4}}e^{\lambda\zslash_2}e^{\zslash_1d}\mu_+
-e^{-i\frac{\tau}{4}}e^{-\lambda\zslash_2}e^{-\zslash_1d}\mu_-\right\},\\ 
\Pi&=\frac{1}{2g}\left(e^{i\frac{\tau}{4}}e^{-\lambda\zslash_2}(y+\yslash)e^{\zslash_1 d}\mu_++e^{-i\frac{\tau}{4}} e^{\lambda\zslash_2}(y-\yslash)e^{-\zslash_1 d} \mu_-  \right).
\end{align}
\newpage

\bibliographystyle{unstr}

\end{document}